\documentclass[aps,pra,eqsecnum,twocolumn]{revtex4}
\usepackage{graphicx}
\usepackage{bm}
\usepackage{amsmath,amssymb,amsthm,dsfont,bm}
\usepackage{color}
\usepackage{wasysym}
\usepackage{ulem}
\usepackage[applemac]{inputenc}

\begin{document}
\preprint{}
\title{Violation of Bell quantum probability inequalities with classical fields}
\author{Laura Ares}
\email{laurares@ucm.es}
\author{Alfredo Luis}
\email{alluis@fis.ucm.es}
\homepage{http://www.ucm.es/info/gioq}
\affiliation{Departamento de \'{O}ptica, Facultad de Ciencias
F\'{\i}sicas, Universidad Complutense, 28040 Madrid, Spain}
\date{\today}

\begin{abstract}
Violations of Bell inequalities in classical optics have been demonstrated in terms of field mean intensities and correlations, however, the quantum meaning of violations point to statistics and probabilities. We present a violation of Bell inequalities for classical fields in terms of probabilities, where we convert classical-field intensities into probabilities via the standard photon-counting equation. We find violation for both, entangled and separable field states. We conclude that any obtained quantum effect might be fully ascribed to the quantum nature of the detector rather than the field itself. Finally, we develop a new Bell-like criterion which is satisfied by factorized states and it is not by the entangled state.
\end{abstract}
\maketitle

\section{Introduction}

There are many physical phenomena originally introduced in the area of quantum physics that were later also found in the realm of classical physics, specially in classical optics. This includes entanglement, complementarity and Bell inequalities as once considered exclusive hallmarks of the quantum theory \cite{AL09,ClassBell,PiO,GBAL18}.

In this regard, proposed violations of Bell inequalities in classical optics are demonstrated in terms of field mean intensities and correlations \cite{ClassBell,PiO}, while the quantum violations point to statistics and probabilities.

\bigskip

In this work we present a violation of Bell inequalities for classical fields in terms of probabilities. The key point is that we convert classical-field intensities into probabilities via the standard photon-counting equation, giving the number of ejected electrons in a photoelectric detection \cite{JP71}.

We focus on genuine quantum probabilities, since the photoelectric effect is a true quantum effect depending on the Planck constant. So we deal with a violation of Bell inequalities for joint probabilities in the quantum realm. However, since the photoelectric effect admits a semiclassical explanation \cite{Photoelectric}, this quantum effect might be fully ascribed to the quantum nature of the detection processes rather than the field itself \cite{LA17}.

We analyze two different situations, these are, the initial state being entangled or factorized. In both cases we find violation of standard Bell inequalities in terms of probabilities. In addition, we study the limit when the intensities are low, so that the probabilities becomes proportional to the intensities and we recover the expected behaviour: violation for the entangled state, no violation for the separable state. Then, we return to the full probabilities and develop a relationship between probabilities which is satisfied by the separable state and it is not by the entangled state.

\section{Scheme} 

\begin{figure}[h]
    \centering
    \includegraphics[width=8cm]{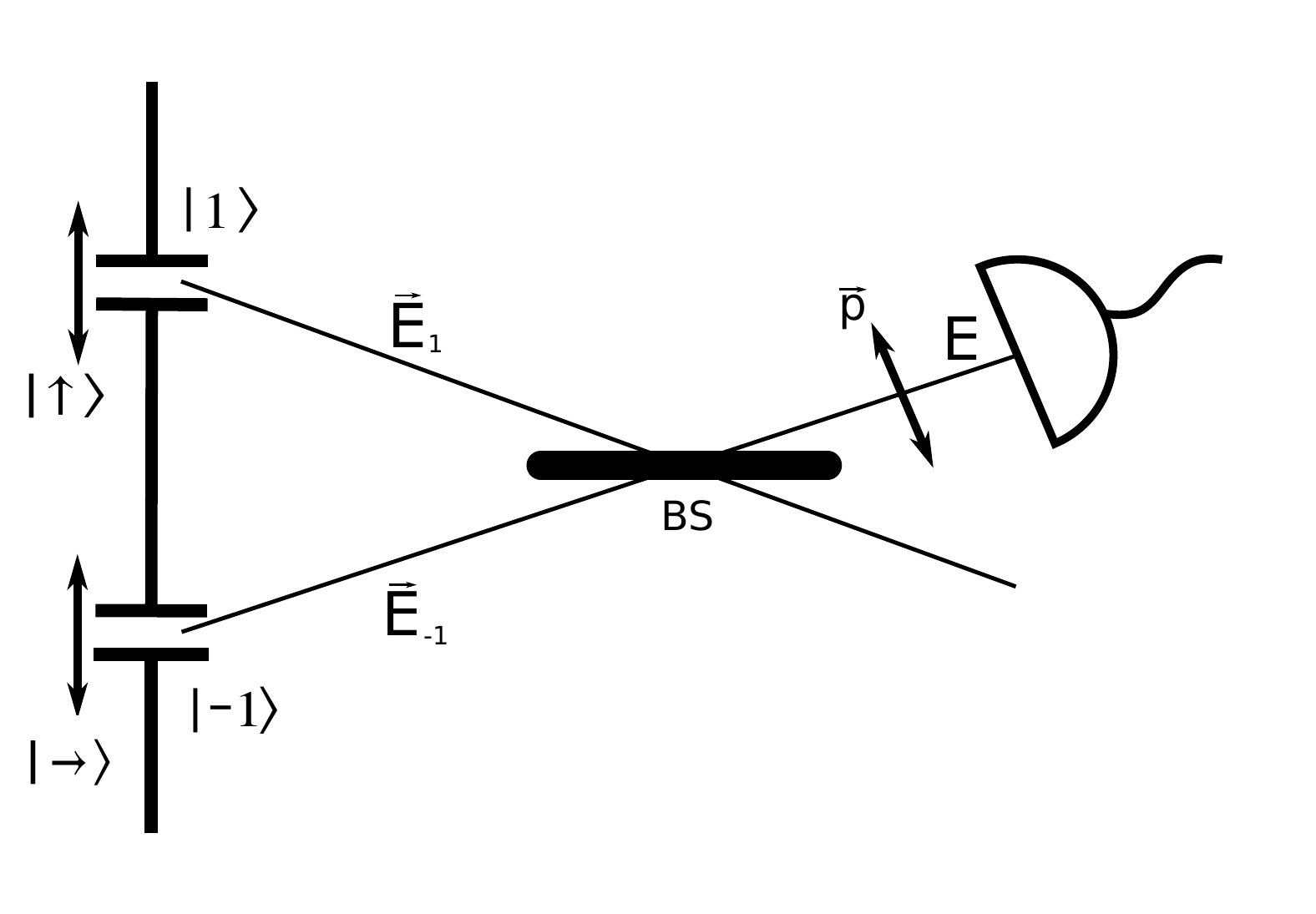}
    \caption{System and observation scheme.}
    \label{Scheme}
\end{figure}

We consider a Young two-beam interference setting with polarized light illustrated in Fig. \ref{Scheme}. The two beams at the apertures are treated as classical fields with different polarization state. This may lead to an entanglement scenario between spacial mode and polarization as the two subsystems to be entangled.

\subsection{Classical entangled field state} 

We begin with the case when the incident beams have orthogonal polarization, say linear vertical $| \uparrow \rangle$ and linear horizontal $ | \rightarrow \rangle$. 
Let us present different notations for the two interfering beams for the sake of clarity and physical insight: 
\begin{eqnarray}
\label{fe}
  &  \bm{E}_1 = E_1 \chi_1 | \uparrow \rangle = \chi_1 \left ( \begin{array}{c}
        E_1 \\ 0 \end{array} \right )= E_1 | 1 \rangle | \uparrow \rangle, & \nonumber \\
   & \\
 &  \bm{E}_{-1} = E_{-1} \chi_{-1} | \rightarrow \rangle =  \chi_{-1} \left ( \begin{array}{c}
        0 \\ E_{-1} \end{array} \right )= E_{-1} | -1 \rangle | \rightarrow \rangle , & \nonumber   
\end{eqnarray}
where the orthogonal, unit vectors 
\begin{equation}
 | \uparrow \rangle = \left ( \begin{array}{c}
        1 \\ 0 \end{array} \right ), \quad | \rightarrow \rangle =   \left ( \begin{array}{c}
        0 \\ 1 \end{array} \right ) ,
\end{equation}
provide a suitable basis for the polarization Hilbert space, while $\chi_{\pm 1}$ are some characteristic functions, depending on the space coordinates $\bm{r}$, expressing the aperture, or say field mode. This can be also expressed in a more quantum fashion terms of the  vectors $|\pm 1\rangle$, say 

\begin{equation}
  \chi_{\pm 1} (\bm{r} ) = \langle \bm{r} | \pm 1 \rangle .
\end{equation}
In the limit of narrow enough apertures the spatial degree of freedom becomes a subsystem with his own two-dimensional Hilbert space spanned by the vectors $|\pm 1 \rangle$, or the functions $\chi_{\pm 1}$. For definiteness we consider always that both field amplitudes $E_{\pm 1}$ are nonfluctuating.

Thus, although the light fields are purely classical, we can nevertheless express it in a quantum-like fashion as the pure state
\begin{equation}
    \bm{E} =  \bm{E}_1 + \bm{E}_{-1} =  E_1  | 1 \rangle | \uparrow \rangle + E_{-1} | -1 \rangle | \rightarrow \rangle =|\psi\rangle.
    \label{ES}
\end{equation}

\bigskip

This is a {\it bona fide} entangled state, involving two well-defined and independent subsystems, space and polarization, with their own Hilbert spaces. It can be argued that the measurement performed is not of the space-type, since space and polarization can not be physically split and spatially separated. This can be also referred to as {\it intra}system entanglement \cite{PiO}, and thus void of nonlocal issues.
 
However, note that with our definition of subsystem the measurements to be performed are strictly local so there is no cross influence of one measurement into the other since they take place in different Hilbert spaces. This to say that the measurement factorizes in the product of independent projectors on the space and polarization spaces, so it satisfies the abstract factorization typical of the derivation of Bell inequalities.

In this regard, we recall that there are many analyses that stress that the key point of Bell inequalities is not nonlocality, but either factorization of statistics \cite{factorization} or lack of a joint distribution including all observables \cite{joint}. Thus, according to Fine's theorem the Bell violation result we pursue would imply that there is no common statistics to all measurements performed, that is really a worth result independent of any claim of locality.

\bigskip
 
\section{Measurement} 

Bell-like scenarios require to perform simultaneous measurements both in the space and polarization subsystems. In our classical scenario this is carried out by properly mixing the fields coming by the apertures, for example via an unpolarizing beam splitter with transmittance and reflectance $\sin^2 \alpha$ and $\cos^2 \alpha$, followed then by a perfect polarizer with axis forming and angle $\beta$ with the horizontal axis as shown in Fig. \ref{Q}. By perfect polarizer we mean null transmittance orthogonal to its axis and unit transmittance along its axis.

The combination of the beam splitter and polarizer can be then expressed in the quantum-like fashion as 
\begin{equation}
\label{Eps}
    E = \langle p | \langle s | \psi \rangle ,
\end{equation}
where $| s \rangle$ and $| p  \rangle$ 

\begin{equation}
     | s \rangle = \bm{s} = \cos \alpha | 1 \rangle + \sin \alpha | -1 \rangle  
\end{equation}
\begin{equation}
     | p \rangle = \bm{p} = \cos \beta | \uparrow \rangle + \sin \beta | \rightarrow \rangle  
\end{equation}
are real vectors fully specifying the measurement performed in the spatial and polarization subsystems.

\bigskip

The vector amplitude of the field incident on the polarizer is 
\begin{equation}
\label{bE}
     \bm{s}\cdot\bm{E}=\cos \alpha E_1 | \uparrow \rangle + \sin \alpha E_{-1} | \rightarrow \rangle , 
\end{equation}
and the amplitude of the field emerging from the polarizer is 
\begin{equation}
    E = \bm{p} \cdot  \left ( \cos \alpha E_1 | \uparrow \rangle + \sin \alpha E_{-1} | \rightarrow \rangle \right).
\end{equation}

\bigskip

So, $E$ is the amplitude of the field illuminating a photoelectric-type quantum detector with some quantum efficiency $\eta$. In the spirit of the typical Bell tests we consider just two only outcomes for the measurement, click and no click, this is no signal versus any signal. The probability that the detector fires  when illuminated with a field of complex amplitude $E$ is 
\begin{equation}
    P = 1 - e^{- \eta T | E|^2} ,
\end{equation}
where $T$ is the detection time \cite{LM59}. 

\bigskip

Finally, for the Bell inequalities we will need also the click probability when there is no polarization or space measurements. For example, when there is no polarizer the click probability is

\begin{equation}
    P_s = 1 - e^{- \eta T | \bm{E} \cdot \bm{s}|^2} ,
\end{equation}
 and likewise 
\begin{equation}
    P_p = 1 - e^{- \eta T | \bm{E} \cdot \bm{p}|^2} ,
\end{equation}
being 
\begin{equation}
    \bm{E} = \left ( \begin{array}{c}
        E_1 \\ E_{-1} \end{array} \right ) .
\end{equation}

\bigskip
\bigskip

\section{Bell inequality} 

With all this we have all the ingredients to compute the Bell-CH inequality for click probabilities \cite{CH74,AF82}, such as 
\begin{equation}
\label{Ineq1}
 0 \geq C  \geq -1, 
\end{equation} where 
\begin{equation}
\label{Ineq2}
 C = P_{x,u} - P_{x,v} + P_{y,u} + P_{y,v}
- P_y  - P_u , 
\end{equation} 
where the subscripts $x,y,u,v$ refer to measurements with two orientations of $\bm{s}$ in the space subsystem ($\bm{s_x}, \bm{s_y}$), and two orientations of $\bm{p}$ in the polarization subsystem ($\bm{p_u}, \bm{p_v}$).

\bigskip

\subsection{Maximally entangled state}

As field state we will consider the maximally entangled state that holds in Eq. (\ref{fe}) for $E_1 = E_{-1}= \varepsilon $. In such a case after Eq. (\ref{Eps})  we get  
\begin{equation}
\label{fa}
    E_{p,s} =  \langle p | \langle s | \psi \rangle = \varepsilon  \bm{p} \cdot \bm{s} .
\end{equation}
and then 
\begin{equation}
    P_{p,s} = 1 - e^{- \kappa \left ( \bm{p} \cdot \bm{s} \right )^2} 
\end{equation}
as well as 
\begin{equation}
    | \bm{E} \cdot \bm{s}|^2 = | \bm{E} \cdot \bm{p}|^2 = |\varepsilon |^2,
\end{equation}
so that
\begin{equation}
    P_p = P_s = 1 - e^{- \kappa} ,
\end{equation}
where $\kappa = \eta T |\varepsilon |^2$.
\color{black}

\bigskip
We have computed $C$ in Eq. (\ref{Ineq2}) as a function of $\kappa$, considering the following particular case for the measurement vectors:
\begin{eqnarray}
\label{Bs}
   & \bm{s}_x = \frac{1}{\sqrt{2}}(1,1), \quad \bm{s}_y = (0,1), & \nonumber \\ & & \\
   & \bm{p}_u = (\cos{\frac{\pi}{3}}
   ,\sin{\frac{\pi}{3}
  }), \quad \bm{p}_v = \frac{1}{\sqrt{2}}(1,-1) .  \nonumber 
\end{eqnarray}

\bigskip

We found a violation of the Bell inequality in  Eq. (\ref{Ineq1}) for all $\kappa \neq 0$, as it can be seen in Fig. 2.

\begin{figure}[h]
    \centering
    \includegraphics[width=8cm]{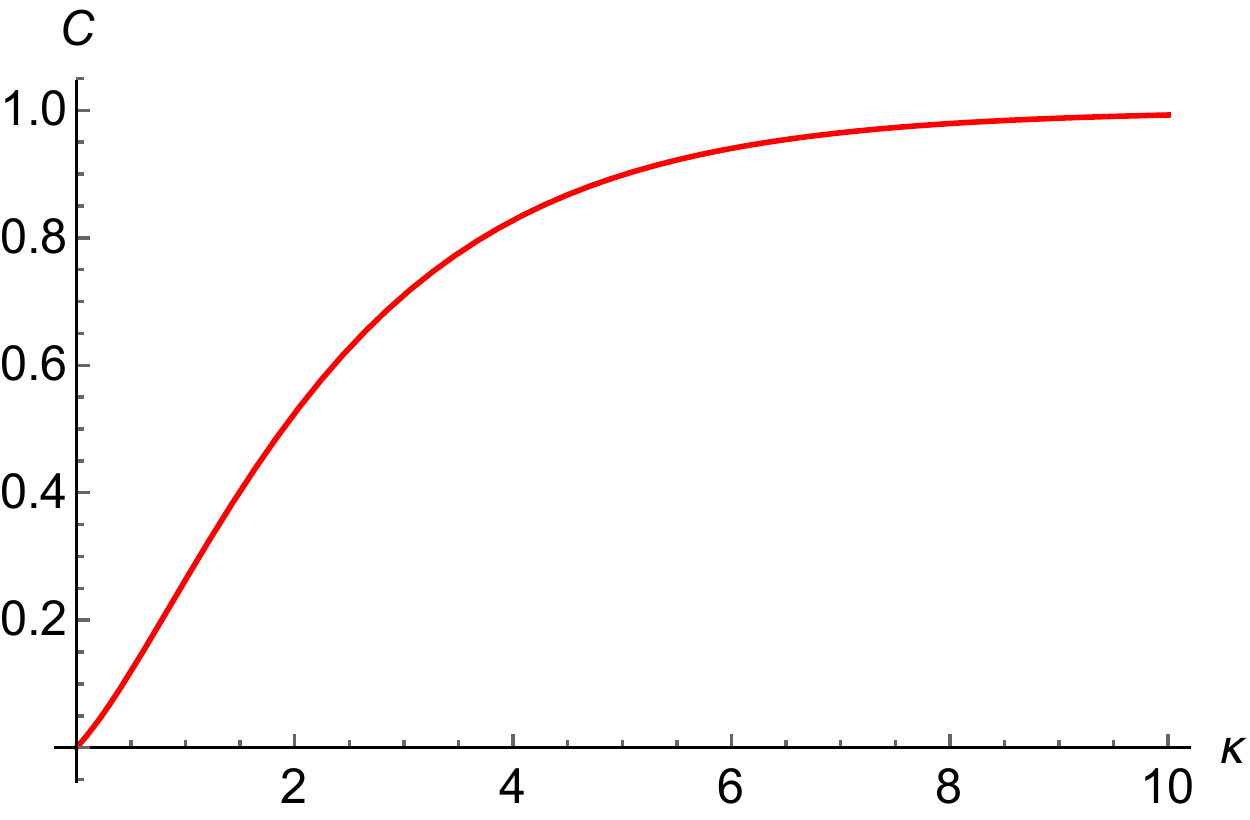}
    \caption{Plot of $C$ for the entangled state showing violation of inequality (\ref{Ineq1}) for all $\kappa \neq 0$.}
    \label{Q}
\end{figure}

\bigskip
\bigskip

\subsection{Separable state}
Let us examine the intriguing possibility of Bell-inequality  violation for separable fields, this is, pure states where space and polarization factorize.  That implies the polarization state being the same at both apertures, say 
\begin{equation}
\label{sf}
    \bm{E} =  \bm{E}_1 + \bm{E}_{-1} = \varepsilon | s_0 \rangle |p_0 \rangle ,
\end{equation}
with 
\begin{eqnarray}
    & | p_0 \rangle = \cos \beta_0 | \uparrow \rangle + \sin \beta_0 | \rightarrow \rangle \equiv \bm{p}_0 , & \nonumber \\ 
     & & \\
& | s_0 \rangle = \cos \alpha_0 | 1 \rangle + \sin \alpha_0 | -1 \rangle \equiv \bm{s}_0 , & \nonumber 
\end{eqnarray}
where the vectors $\bm{p}_0, \bm{s}_0$ are real vectors specifying the space and polarization state of the field emerging from the apertures. We have chosen them real for the sake of simplicity. 

In this case the vector amplitude of the field incident on the polarizer is 
\begin{equation}
\label{bEs}
    \bm{E} = \varepsilon \bm{s} \cdot \bm{s}_0  | p_0 \rangle , 
\end{equation}
so the amplitude of the field emerging from the polarizer is 
\begin{equation}
    E =  \varepsilon  \left ( \bm{s} \cdot \bm{s}_0 \right ) \left ( \bm{p} \cdot \bm{p}_0 \right ) ,
\end{equation}
and then 
\begin{equation}
    P_{p,s} = 1 - e^{- \kappa \left ( \bm{s} \cdot \bm{s}_0 \right )^2 \left ( \bm{p} \cdot \bm{p}_0 \right )^2} .
    \label{PpsSB}
\end{equation}
Finally, the click probability when there is no polarizer  depends on the modulus of the electric field leaving the upper arm of the beam splitter in Eq. (\ref{bEs})
\begin{equation}
    P_s = 1 - e^{- \eta T | \bm{E}|^2} = 1 - e^{- \kappa \left ( \bm{s} \cdot \bm{s}_0 \right )^2 },
    \label{PsSB}
\end{equation}
for the same $\kappa = \eta T |\varepsilon |^2$ as before. Likewise, if there were no beam splitter the click probability would be given by the Malus' law applied to the field at the apertures (\ref{sf}) 
\begin{equation}
    P_p = 1 - e^{- \kappa\left ( \bm{p} \cdot \bm{p}_0 \right )^2}.
    \label{PpSB}
\end{equation}

\bigskip
\bigskip

\begin{figure}[h]
    \centering
    \includegraphics[width=8
cm]{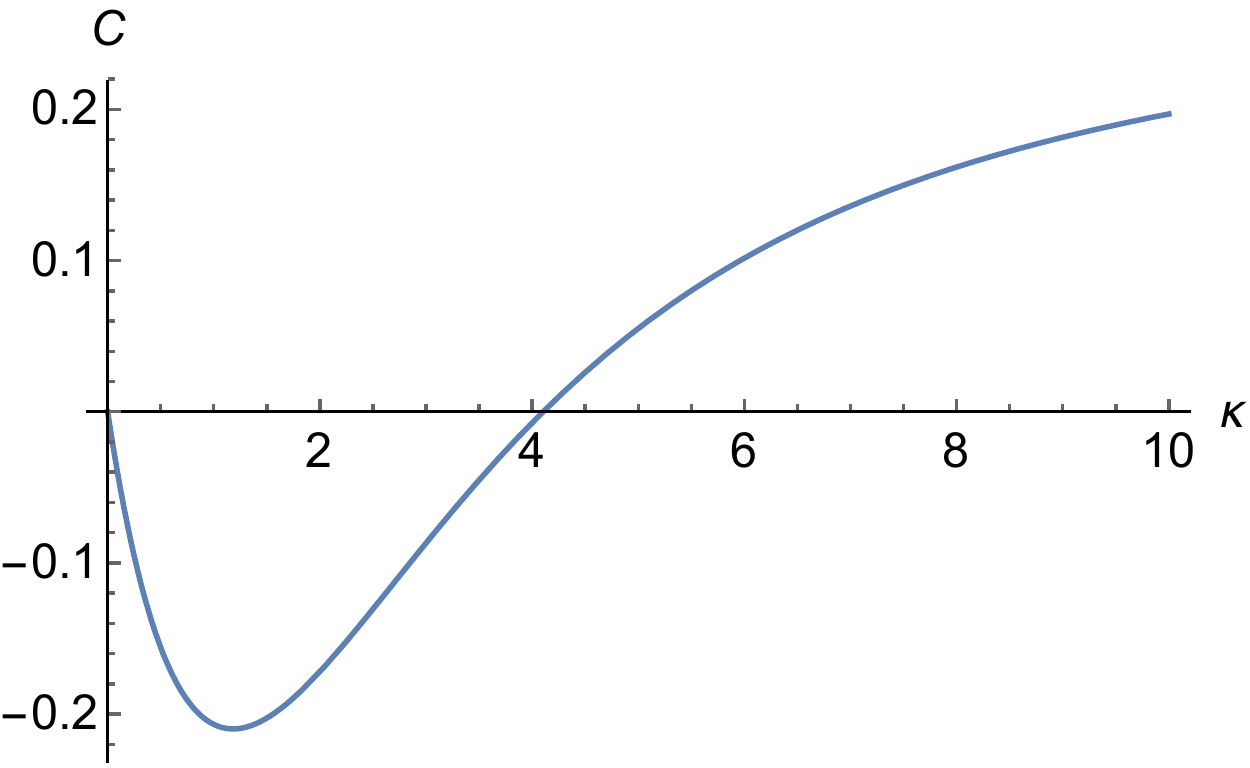}
    \caption{Plot of $C$, for separable factorized states with $\alpha_0=\pi/2$ and $\beta_0=\pi/3$.
     .}
    \label{SB}
\end{figure}

In Fig. \ref{SB} we show an example of how separable states also violate inequality (\ref{Ineq1}) for the same measurement settings on Eq. (\ref{Bs}).
\bigskip

The reason for this Bell-inequality violation for separable fields lies in the fact that the relation between click probabilities and modulus square of the field is highly nonlinear. This is at sharp contrast with quantum mechanics in which this relation is linear almost by definition.

\begin{figure}[h]
    \centering
    \includegraphics[width=8.5cm]{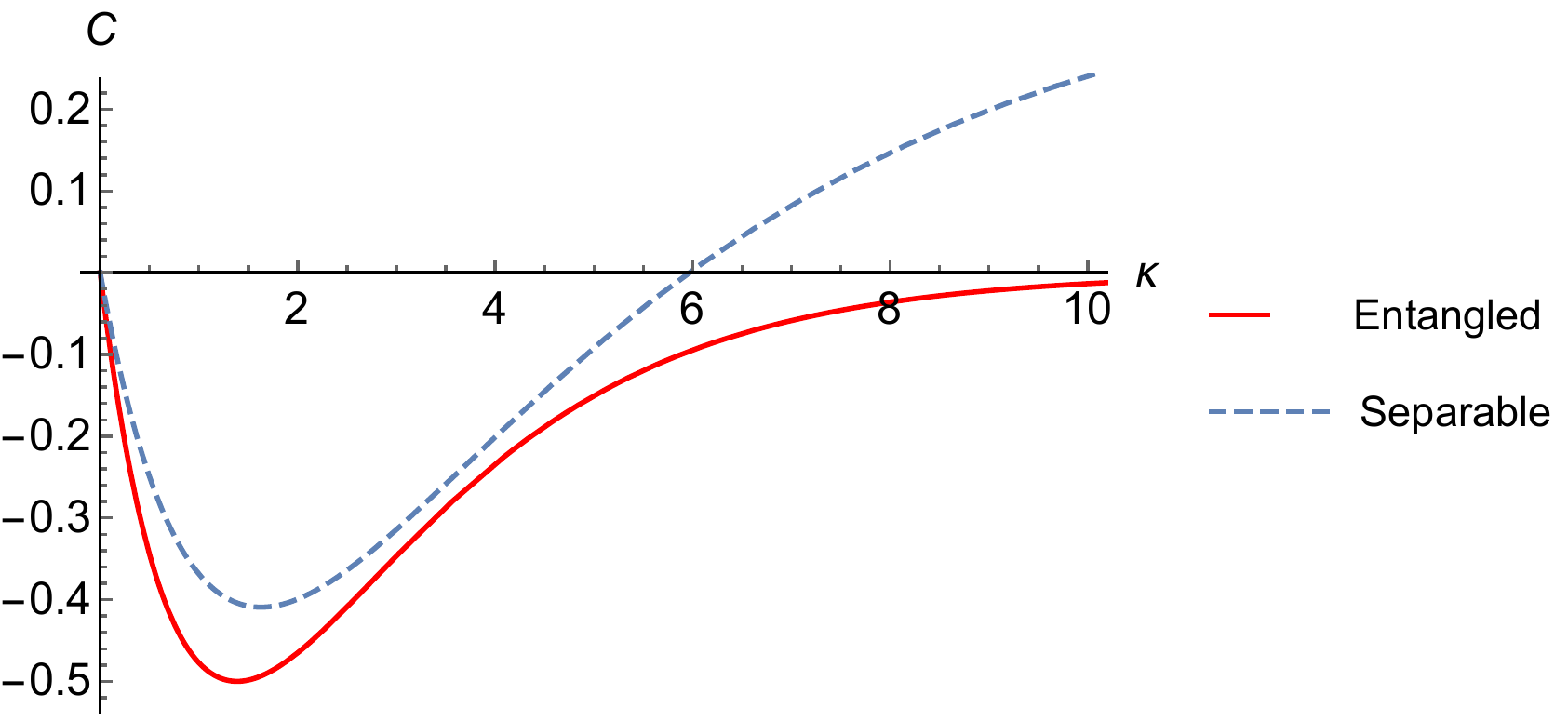}
    \caption{Plot of $C$, for the maximally entangled state and separable state with $\alpha_0=\pi/3$ and $\beta_0=\pi/8$.}
     
    \label{CBSB}
\end{figure}

Moreover, it can be found measurements for which separable states violate Bell inequality (\ref{Ineq1}), while the entangled state does not. This occurs, for example, in the case of
\begin{eqnarray}
   & \bm{s}_x = (1,0), \quad \bm{s}_y = (0,1), & \nonumber \\ & & \\
   & \bm{p}_u = \frac{1}{\sqrt{2}}(1,1) 
  , \quad \bm{p}_v = \frac{1}{\sqrt{2}}(1,-1),  \nonumber 
\end{eqnarray}

as illustrated in Fig \ref{CBSB}.

\bigskip

This further suggests two points worth checking: recovering a linear relation between probabilities and intensities, and derive new nonlinear Bell-type click inequalities adapted to witness entanglement in this scenario.

\bigskip
\subsection{Limit of vanishing intensities}

We analyze the case of $ \kappa \rightarrow 0$ so that the photo-count probabilities admits a linear approximation. In this limit we recover the standard idea that click probability is proportional to field intensity. So we can approach the case of Bell violation in terms of classical field intensities at the same time that we approach the quantum regime of a single photon. Intuitively this limit be checked in Figs \ref{Q} and \ref{SB} showing that in the separable factorized case we get into the region of no Bell-violation $C<0$, while the entangled state violates the inequality (\ref{Ineq2}) always, even in the limit of vanishing intensities.

Next, we show that separable states do not violate Bell inequalities in the limit of low intensities. To this end, it is worth considering general measurement settings in the form 
\begin{eqnarray}
\label{settings}
   &\bm{s}_x = (\cos x ,\sin x), \quad \bm{s}_y = (\cos y ,\sin y), & \nonumber \\ & & \\
   & \bm{p}_u = (\cos u ,\sin u), \quad \bm{p}_v = (\cos v ,\sin v) . & \nonumber 
\end{eqnarray}
In the limit of vanishing intensities, the click probabilities (\ref{PpsSB})-(\ref{PpSB}) can be approximated by its first term of the Taylor series, so that
\begin{widetext}
\begin{equation}
\small
 C= \kappa\left[ \left ( \bm{s}_x \cdot \bm{s}_0 \right )^2 \left ( \bm{p}_v \cdot \bm{p}_0 \right )^2+ \left ( \bm{s}_y \cdot \bm{s}_0 \right )^2+ \left ( \bm{p}_v \cdot \bm{p}_0 \right )^2- \left ( \bm{s}_x \cdot \bm{s}_0 \right )^2 \left ( \bm{p}_u \cdot \bm{p}_0 \right )^2- \left ( \bm{s}_y \cdot \bm{s}_0 \right )^2 \left ( \bm{p}_u \cdot \bm{p}_0 \right )^2-  \left ( \bm{s}_y \cdot \bm{s}_0 \right )^2 \left ( \bm{p}_v \cdot \bm{p}_0 \right )^2\right]
\end{equation}
\begin{equation}
=\dfrac{\kappa}{2} \left\{-1+\cos \left[2(\alpha_0-x)\right] \sin(u-v) \sin(2 \beta_0-u-v) \nonumber+ \cos[2(\alpha_0-y)] \cos(u-v) \cos(2 \beta_0-u-v)  \right\} .
\label{CSlin}
\end{equation}
It can be seen that there is not violation of inequality (\ref{Ineq2}) by checking its higher and lower values:
\begin{eqnarray}
&C\leq\dfrac{\kappa}{2} \left[ -1+ \sin\left(u-v\right) \sin\left(2 \beta_0-u-v\right)+ \cos \left(u-v\right) \cos\left(2 \beta_0-u-v\right) \right]  \nonumber
& \\ \nonumber\\
&=\dfrac{\kappa}{2}\left\{-1+\cos[2(\beta_0-u)]\right\}\leq 0,
\label{LClin}
\end{eqnarray}
and
\begin{eqnarray}
&C\geq\dfrac{\kappa}{2} \left[-1 -\sin(u-v) \sin(2 \beta_0-u-v) -\cos (u-v) \cos(\beta_0-u-v)\right]  \nonumber
&\\\nonumber\\
&=\dfrac{\kappa}{2}\left\{-1-\cos[2(\beta_0-u)]\right\}\geq-\kappa,
\label{UClin}
\end{eqnarray}

\noindent{where the regime of $\kappa<<1$ always leads to $C\geq -1$.}
\bigskip
\end{widetext}

\section{Nonlinear Bell inequality}
As we have seen, inequality (\ref{Ineq1}) can be violated by both, entangled and separable initial states. We have already commented that this is due to the nonlinear relation between statistics and fields. Thus, this calls for a different form of Bell inequalities able to witness field entanglement taking into account the nonlinearity. Thus, we look for a different relation between probabilities which ideally will only be violated when the initial state is entangled. For definiteness  we will always consider just pure states. 

\bigskip
Let us start from the Bell inequality applied to separable states in the limit of low intensities, (\ref{CSlin}),which we have seen in the previous section that satisfies (\ref{UClin}) and (\ref{LClin}):
\begin{equation}
   0\geq C \geq -\kappa.
\end{equation}

In order to generalize this inequality, capable of distinguishing between separable and no separable states, we develop
\begin{equation}
e^0\geq e^{C}\geq e^{-\kappa},
\end{equation}

\noindent where\\

\noindent $e^C=\frac{e^{-\kappa\left ( \bm{s_x} \cdot \bm{s}_0 \right )^2 \left ( \bm{p_v} \cdot \bm{p}_0 \right )^2}e^{-\kappa\left ( \bm{s_y} \cdot \bm{s}_0 \right )^2}e^{-\kappa\left ( \bm{p_u} \cdot \bm{p}_0 \right )^2}}{e^{-\kappa \left ( \bm{s_x} \cdot \bm{s}_0 \right )^2 \left ( \bm{p_u} \cdot \bm{p}_0 \right )^2}e^{-\kappa\left ( \bm{s_y} \cdot \bm{s}_0 \right )^2 \left ( \bm{p_u} \cdot \bm{p}_0 \right )^2}e^{-\kappa\left ( \bm{s_y} \cdot \bm{s}_0 \right )^2 \left ( \bm{p_v} \cdot \bm{p}_0 \right )^2}}.$\\

\color{black}
\noindent{Remembering that the general probabilities are defined as:}
\begin{eqnarray}
\label{Pp}
&P_{y}=1-e^{-\kappa \left ( \bm{s}_y \cdot \bm{s}_0 \right )^2} ,\nonumber&\\
&P_{v}=1-e^{-\kappa \left ( \bm{p}_u \cdot \bm{p}_0 \right )^2} ,&\\
&P_{ab}=1-e^{-\kappa \left ( \bm{s}_a \cdot \bm{s}_0 \right )^2 \left ( \bm{p}_b \cdot \bm{p}_0 \right )^2},\nonumber&
\end{eqnarray}{}
for $ab=xu,xv,yu,uv$,
we finally arrive to

\begin{equation}
1\geq \mathcal{C} \geq e^{-\kappa}
\label{Ineq3}
\end{equation}{}
with
\begin{equation}
\mathcal{C}= \frac{(1-P_{xv})(1-P_{y})(1-P_{u})}{(1-P_{xu})(1-P_{yu})(1-P_{yv})}.  
\label{Cnonlin}
\end{equation}{}

This is the desired nonlinear Bell-inequality, which is constructed to be satisfied by all pure separable states. In particular, for the example of Fig. \ref{SB} we get
\begin{equation}
    \mathcal{C} = e^{-\kappa/2} ,
\end{equation}
which is well within the bounds (\ref{Ineq3}).

\bigskip

Likewise, let us show that this inequality is violated by the  maximally entangled state (\ref{ES}) for the following measurement settings in Eq. (\ref{settings})  $x=1.7,\; y=1.5,\; u=0, \; v=6.1$, and $\kappa \neq 0$:
\begin{equation}
\mathcal{C} = e^{-2 k} < e^{-\kappa}.  
\end{equation}{}

So, we have derived an inequality which is violated by the entangled state and it is not by the separable state.

\bigskip

\section{Conclusions}

We have presented a semiclassical model of Bell inequalities. We have shown that there are violations of Bell inequalities for quantum probabilities caused by illuminating quantum detectors with classical light.

This result points to the often overlooked effect of measurement on the quantum to classical properties of observed systems. We mean that according to the Born's rule system and measurement play fully symmetric roles, so both contribute to nonclassical results \cite{LA17}. This forgetfulness is quite noticeable within the quantum theory that actually recognizes the importance of measurement in physics.

Moreover we have found clear violations for separable field states. This is possible because the relation between quantum click probabilities and classical field amplitudes is highly nonlinear, at difference with the pure quantum case. 

In this regard we have derived a new nonlinear Bell-type criterion satisfied by all separable, pure field states.

We think all this opens a promising research line combining classical systems with quantum detectors in a nonlinear scenario.      

\bigskip

\noindent{\bf Acknowledgments.- }
L. A. and A. L. acknowledge financial support from Spanish Ministerio de Econom\'ia y Competitividad Project No. FIS2016-75199-P.
L. A. acknowledges financial support from European Social Fund and the Spanish Ministerio de Ciencia Innovaci\'{o}n y Universidades, Contract Grant No. BES-2017-081942.


\bigskip


\bigskip

\begin{thebibliography}{00}

\bibitem{AL09}
A. Luis, Coherence, polarization, and entanglement for classical light fields, Opt. Commun. {\bf 282}, 3665-3670 (2009) 

\bibitem{ClassBell}
P. Suppes, J. A. de Barros, and A. S. Sant'Anna, A Proposed Experiment Showing that Classical Fields Can Violate Bell's Inequalities, arXiv:quant-ph/9606019v1.
R.J.C. Spreeuw, A Classical Analogy of Entanglement, Found. Phys. {\bf 28}, 361--374 (1998);
X.-F. Qian and J. H. Eberly, Entanglement and classical polarization states, Opt. Lett.  {\bf 36}, 4110--4112 (2011);
 K. H. Kagalwala, G. Di Giuseppe, A. F. Abouraddy, and B. E. A. Saleh, Bell's measure in classical optical coherence, Nat. Photon. {\bf 7}, 72--78 (2013);
 F. De Zela, Relationship between the degree of polarization, indistinguishability, and entanglement, Phys. Rev. A {\bf 89}, 013845 (2014); 
 F. T\"{o}ppel, A. Aiello, Ch. Marquardt, E. Giacobino, and G. Leuchs, Classical entanglement in polarization metrology, New J. Phys. {\bf 16}, 073019 (2014);
X.-F. Qian, B. Little, J. C. Howell, and J. H. Eberly, Shifting the quantum-classical boundary: theory and experiment for statistically classical optical fields, Optica {\bf 2}, 611--615 (2015);
F. De Zela, Beyond Bell's theorem: realism and locality without Bell-type
correlations, Scientific Reports {\bf 7}, 14570  (2017);
M. Markiewicz, D. Kaszlikowski, P. Kurzy\'{n}ski, and A. W\'{o}jcik,
From contextuality of a single photon to realism of an electromagnetic wave,
npj Quantum Information {\bf 5}, 5 (2019).

\bibitem{GBAL18}
R. Galazo, I. Bartolom\'e, L. Ares, and A. Luis, Classical and quantum complementarity, Phys. Lett. A
{\bf 384}, 33, 126849 (2020).

\bibitem{PiO}
A. Forbes, A. Aiello, and B. Ndagano, Classically Entangled Light, Progress in Optics, Volume {\bf 64}, 99-153 (2019);
 
\bibitem{JP71}
J. Pe\v{r}ina, Coherence of Light (Van Nostrand Reinhold, 1971);
L. Mandel and E. Wolf, Optical Coherence and Quantum Optics (Cambridge University Press, 1995). 
 
\bibitem{Photoelectric}
W. E. Lamb and M. O. Scully, The photoelectric effect without photons, in {\it Polarization, Matter and Radiation},  Volume in Honour of A. Kastler (Presses Universitaires de France, Paris, 1969);
A. Muthukrishnan, M. O. Scully, and M.  Zubairy, The concept of the photon - Revisited, Optics and Photonics News. \textbf{14}, 18-27 (2003). 
G. Greenstein, A. G. Zajonc, The Quantum Challenge: Modern Research on the Foundations of Quantum Mechanics, Jones \& Bartlett Learning, (1997).



\bibitem{LA17}
A. Luis and L. Ares, Apparatus contribution to observed nonclassicality, Phys. Rev. A {\bf 102}, 022222 (2020).

\bibitem{factorization}
A. Fine, Correlations and Physical Locality, PSA: Proceedings of the Biennial Meeting of the Philosophy of Science Association, Vol. 1980, Volume Two: Symposia and Invited Papers (1980), pp. 535-562; 
A. Fine, Antinomies of Entanglement: The Puzzling Case of the Tangled Statistics, The Journal of Philosophy, {\bf 79}, 733-747 (1982).

\bibitem{joint}
N. N. Vorob'ev, Consistent families of measures and their extensions, Theor. Probab. Applications {\bf VII}, 147 (1962);
M. Czachor, On some class of random variables leading to violations of the Bell inequality, Phys. Lett. A {\bf 129}, 291--294 (1988); Erratum, Phys. Lett. A {\bf 134}, 512(E) (1989);
A. Fine, Hidden Variables, Joint Probability, and the Bell Inequalities, Phys. Rev. Lett {\bf 48}, 291--295 (1982);
A. Fine, Joint distributions, quantum correlations, and commuting observables
J. Math. Phys. {\bf 23}, 1306--1310 (1982);
A. Matzkin, Is Bell's theorem relevant to quantum mechanics. On locality and non-commuting observables, arXiv: 0802.0613 [quant-ph];
A. Khrennikov, Non-Kolmogorov probability models and modified Bell's inequality, arXiv:0003.017 [quant-ph];
J. Christian, On a Surprising Oversight by John S. Bell in the Proof of his Famous Theorem, arXiv:1704.02876 [physics.gen-ph];
A. Khrennikov, CHSH inequality: Quantum probabilities as classical conditional probabilities, arXiv:1406.4886 [quant-ph];
K. Hess and W. Philipp, Bell's theorem: Critique of proofs with and without inequalities,  arXiv: 0410.015 [quant-ph];
T. M. Nieuwenhuizen, Is the Contextuality Loophole Fatal for the Derivation of Bell Inequalities?, Found. Phys. {\bf 41}, 580 (2011).
E. Masa, L. Ares, and A. Luis, Nonclassical joint distributions and Bell measurements, Phys. Lett. A {\bf 384}, 126416 (2020).

\bibitem{LM59}
L. Mandel, Fluctuations of Photon Beams: The Distribution of the Photo-Electrons, Proc. Phys. Soc. {\bf 74}, 233 (1959).

\bibitem{CH74}
J.F. Clauser, M.A. Horne, Experimental consequences of objective local theories,
Phys. Rev. D {\bf 10}, 526-535 (1974).

\bibitem{AF82}
A. Fine, Hidden variables, joint probability, and the Bell inequalities, Phys. Rev.
Lett. {\bf 48},291-295 (1982).

\bibitem{SL18}
C. Sanchidri\'an Vaca and A. Luis, Entanglement between total intensity and polarization for pairs of coherent states, Phys. Rev. A {\bf 97}, 043810 (2018).



\end{thebibliography}
\end{document}